\documentclass{appolb}
\usepackage{epsfig}

\begin{document}

\title{LANGEVIN APPROACH TO L\'{E}VY FLIGHTS IN FIXED
POTENTIALS: EXACT RESULTS FOR STATIONARY PROBABILITY DISTRIBUTIONS
\thanks{Presented at the XIX Marian
Smoluchowski Symposium on Statistical Physics, \ \newline Krakow,
Poland, May 14-17, 2006.}}
\author{Alexander Dubkov\footnote{e-mail: dubkov@rf.unn.ru}
\address{Radiophysics Department, Nizhni Novgorod State
University \ \\23 Gagarin Ave., 603950 Nizhni Novgorod, Russia}
\and Bernardo Spagnolo
\address{Dipartimento di Fisica e Tecnologie Relative, Universit\`a di
Palermo \ \\CNISM - Unit\`a di Palermo, Group of Interdisciplinary
Physics\footnote{http://gip.dft.unipa.it} \ \\Viale delle Scienze,
I-90128 Palermo, Italy} } \maketitle

\begin{abstract}
The functional method to derive the fractional Fokker-Planck
equation for probability distribution from the Langevin equation
with L\'{e}vy stable noise is proposed. For the Cauchy stable noise
we obtain the exact stationary probability density function of
L\'{e}vy flights in different smooth potential profiles. We find
confinement of the particle in the superdiffusion motion with a
bimodal stationary distribution for all the anharmonic symmetric
monostable potentials investigated. The stationary probability
density functions show power-law tails, which ensure finiteness of
the variance. By reviewing recent results on these statistical
characteristics, the peculiarities of L\'{e}vy flights in comparison
with ordinary Brownian motion are discussed.
\end{abstract}


\section{Introduction}

Anomalous diffusion in the form of L\'{e}vy flights appears in
many physical, chemical, biological, and financial
systems~\cite{Met00}--\cite{Bar04}. L\'{e}vy flights are
stochastic processes characterized by the occurrence of extremely
long jumps. The length of these jumps is distributed according to
a L\'{e}vy stable statistics with a power law tail and divergence
of the second moment. This peculiar property strongly contradicts
the ordinary Brownian motion, for which all moments of the
particle coordinate are finite.  The presence of anomalous
diffusion can be explained as a deviation of real statistics of
fluctuations from Gaussian law, that it has lead to the
generalization of the central limit theorem by L\'{e}vy and
Gnedenko~\cite{Lev25}--\cite{Gne54}. The divergence of the
variance of L\'{e}vy flights poses some problems as regards to the
physical meaning of these processes. However, recently the
relevance of L\'{e}vy motions appeared in many physical, natural
and social complex systems. The L\'{e}vy type statistics, in fact,
is observed in various scientific areas. Among many interesting
examples we cite here the subrecoil laser
cooling~\cite{Bar94}--\cite{Sch99}, the diffusion by flows in
porous media~\cite{Pai96}, the fluctuations in
plasmas~\cite{Che02b}, the molecular collisions~\cite{Car03}, the
spatial gazing patterns of bacteria~\cite{Lev97}, the flights of
an albatross~\cite{Vis96}, the long paleoclimatic time series of
the Greenland ice core measurements~\cite{Ditl99}, and the
financial time series~\cite{Man63}--\cite{Man96}. Experimental
evidence of L\'{e}vy processes was also observed in the motion of
single ion in a one-dimensional optical lattice~\cite{Kat97} and
in the particle evolution along polymer chains~\cite{Sok97,Lom05}.

The problem of the barrier crossing in a bistable potential, the
particle escape from a metastable state, and the first passage
time density have been analyzed, recently, for L\'{e}vy
flights~\cite{Dit99}--\cite{Kor07}. The main focus in these papers
is to understand how the barrier crossing behavior, according to
the Kramers law~\cite{Kra40}, is modified by the presence of the
L\'{e}vy noise.

L\'{e}vy flights are a special class of Markovian processes,
therefore the powerful methods of the Markovian analysis are in
force in this case. We mean a possibility to investigate the
stationary probability distributions of superdiffusion, the first
passage time and the residence time characteristics, the spectral
characteristics of stationary motion, etc. Of course, this type of
diffusion has a lot of peculiarities different from those observed
in normal Brownian motion. The main difference from ordinary
diffusion consists in replacing the white Gaussian noise source in
the underlieing Langevin equation with a L\'{e}vy stable noise.

In this paper we use functional approach to derive the
Fokker-Planck equation, with fractional space derivative, directly
from Langevin equation with a L\'{e}vy stable noise source.
Starting from this equation we find the exact stationary
probability distribution (SPD) of fast diffusion in symmetric
smooth monostable potentials for the case of Cauchy stable noise.
Specifically, we consider symmetric potential profiles $U(x) =
\gamma x^{2m}/(2m)$ (with odd $m=2n + 1$ and even $m = 2n$),
describing the dynamics of overdamped anharmonic oscillator driven
by L\'{e}vy noise. We find that for L\'{e}vy flights in steep
potential well, with steepness greater or equal to four, the
variance of the particle coordinate is finite. This gives rise to
a confined superdiffused motion, characterized by a bimodal
stationary probability density, as previously reported in
Refs.~\cite{Che04}--\cite{Che02}. However, in previous
Ref.~\cite{Che02} the authors analyzed the properties of
stationary probability distribution for nonlinear L\'{e}vy
oscillators and its bimodality as a function of the L\'{e}vy index
$\alpha$, by finding that it is more pronounced for $\alpha = 1$
(Cauchy stable noise) and becomes the Boltzmann stationary
distribution in the limit of $\alpha \rightarrow 2$. Here we
analyze the SPDs as a function of a dimensionless parameter
$\beta$, which is the ratio between the noise intensity $D$ and
the steepness of the potential profile $\gamma$. We find that the
SPDs remain bimodal with increasing $\beta$ parameter, that is
with decreasing the steepness $\gamma$ of the potential profile,
or by increasing the noise intensity $D$.

\section{Functional method to derive the fractional Fokker-Planck equation
from Langevin equation with L\'{e}vy stable noise}

Ditlevsen and Yanovsky with co-authors for the first time obtained
the fractional Fokker-Planck equation directly from Langevin
equation, by replacing the white Gaussian noise with L\'{e}vy
stable noise~\cite{Dit99,Sch01} (see also~\cite{Yan00,Gar00}).
However, some attempts were undertaken before in
Ref.~\cite{Fog94,Jes99}.

The theory of L\'{e}vy processes is closely linked to that of
infinitely divisible distributions~\cite{Fel71}-\cite{Uch99}.
Therefore, starting from this link, we have recently developed a
more general approach, based on the theory of infinitely divisible
distributions and functional analysis, to derive the generalized
Kolmogorov equation for arbitrary non-Gaussian white noise
source~\cite{Dub05}. Here we obtain the equation for probability
distribution from Langevin equation with L\'{e}vy stable noise, by
a different approach with respect to that reported in
Ref.~\cite{Sch01}.

Let us consider the anomalous overdamped motion in the potential
profile $U\left( x\right)$
\begin{equation}
\frac{dx}{dt}= - \;U^{\prime }(x)+L(t).  \label{Lang}
\end{equation}
Here $x(t)$ is the displacement of particle and $L\left( t\right)$
is the symmetric $\alpha $-stable L\'{e}vy noise with the
characteristic function of increments
\begin{eqnarray}
\left\langle \exp \left\{ ik\left[ \eta_{L}\left( t+\Delta t
\right) -\eta_{L}\left( t\right) \right] \right\} \right\rangle
&=& \left\langle \exp \left\{ ik\int\limits_{t}^{t+\Delta
t}L\left( \tau
\right) d\tau \right\} \right\rangle \nonumber \\
&=&\exp \left\{ -D\left\vert k\right\vert ^{\alpha }\Delta t\right\}
, \label{Ch-func}
\end{eqnarray}
where $\eta_L(t)$ is a generalized Wiener process~\cite{Dub05,Str92}
which derivative is the L\'{e}vy stable noise ($d\eta_L /dt =
L(t)$). Here $\alpha $ is the L\'{e}vy exponent $\left( 0<\alpha
<2\right) $ and $D$ is the intensity of L\'{e}vy noise. The case
$\alpha =1$ corresponds to L\'{e}vy noise $ L\left( t\right) $ with
symmetric Cauchy distribution. First of all, we calculate the
characteristic functional of the noise $L\left( t\right) $.

According to the definitions of the characteristic functional of the
random process $L(t)$ and the Stiltjes integral we have
\begin{eqnarray}
\Theta_{t}\left[ u \right] &=& \left\langle \exp\left\{
i\int\limits_{0}^{t}u\left( \tau\right) L\left( \tau\right)
d\tau\right\} \right\rangle = \left\langle \exp\left\{ i
\int\limits_{0}^{t}u\left( \tau\right) d\eta_{L}\left(
\tau\right) \right\} \right\rangle \nonumber \\
&=& \left\langle \exp\left\{
i\lim\limits_{\delta_{\tau}\rightarrow0}
\sum\limits_{k=1}^{n}u\left( \vartheta_{k}\right) \left[
\eta_{L}\left( \tau_{k}\right) -\eta_{L}\left( \tau_{k-1}\right)
\right] \right\} \right\rangle \nonumber \\
&=& \lim\limits_{\delta_{\tau}\rightarrow0}\left\langle
\prod\limits_{k=1} ^{n}\exp\left\{ iu\left( \vartheta_{k}\right)
\left[ \eta_{L}\left( \tau _{k}\right) -\eta_{L}\left(
\tau_{k-1}\right) \right] \right \} \right\rangle , \nonumber
\end{eqnarray}
where $\vartheta _k$ is some internal point of the time interval
$\left( \tau_{k-1},\tau _k\right) \,$, $\delta _\tau =\max
\limits_k\Delta \tau _k$, $ \Delta \tau _k=\tau _k-\tau _{k-1}$
($\tau _0=0$, $\tau _n=t$). Taking into account that the increments
of non-overlapping time intervals of the generalized Wiener process
$\eta_{L}(t)$ are statistically independent and using
Eq.~(\ref{Ch-func}) we obtain
\begin{eqnarray}
\Theta_{t}\left[ u\right] &=&
\lim\limits_{\delta_{\tau}\rightarrow0}\prod\limits_{k=1}
^{n}\left\langle \exp\left\{ iu\left( \vartheta_{k}\right) \left[
\eta_{L}\left( \tau _{k}\right) -\eta_{L}\left( \tau_{k-1}\right)
\right] \right \} \right\rangle \nonumber \\
&=&\lim\limits_{\delta_{\tau}\rightarrow0}\prod\limits_{k=1}
^{n}\exp \left\{ -D\left\vert u\left( \vartheta_{k}\right)
\right\vert ^{\alpha }\Delta \tau_{k}\right\} \\
&=& \exp \left\{
-D\lim\limits_{\delta_{\tau}\rightarrow0}\sum\limits_{k=1}
^{n}\left\vert u\left( \vartheta_{k}\right) \right\vert ^{\alpha
}\Delta \tau_{k}\right\}.\nonumber
\end{eqnarray}
By using the definition of the Riemann integral we finally get
\begin{equation}
\Theta_{t}\left[ u \right] =\exp \left\{
-D\int\limits_{0}^{t}\left\vert u\left( \tau \right) \right\vert
^{\alpha }d\tau \right\}.\label{Funct}
\end{equation}

To derive the fractional Fokker-Planck equation from Langevin
equation~(\ref{Lang}) we need the functional correlational formula
for symmetric $\alpha$--stable L\'{e}vy noise $L\left( t\right) $.
We start from the generalization of Furutsu-Novikov
formula~\cite{Fur63,Nov65} for arbitrary non-Gaussian random process
$\xi(t)$, obtained previously in~\cite{Kly74},
\begin{equation}
\left\langle \xi \left( t\right) R_t\left[ \xi +z\right]
\right\rangle =\left. \frac{\dot \Phi _t\left[ u\right] }{iu\left( t\right) }%
\right| _{\,u=\frac{\delta }{i\delta z}}\left\langle R_t\left[ \xi
+z\right] \right\rangle , \label{Kly}
\end{equation}
where $R_{t}\left[ \xi\right] $ is a functional of noise $\xi \left(
t\right) $, defined on the observation interval $\left( 0,t\right)
$, $z\left( t\right) $ is a deterministic function, and $\Phi
_t\left[ u\right] =\ln \Theta _t\left[ u\right] $. Following
Klyatskin, we use the translation functional operator and taking
into account that the function $z(t)$ is deterministic we have
\begin{equation}
\left\langle \xi \left( t\right) R_t\left[ \xi +z\right]
\right\rangle = \left\langle \xi \left( t\right)
\exp\left\{\int\limits_{0}^{t}\xi \left( \tau \right) \frac{\delta
}{\delta z\left( \tau \right) }d\tau \right\}\right\rangle
R_t\left[ z\right] . \label{F-1}
\end{equation}
For the average entering in Eq.~(\ref{F-1}), after evident
rearrangements, we find
\begin{eqnarray}
\left\langle \xi \left( t\right) \exp\left\{i \int\limits_{0}^{t}
\xi \left( \tau \right) u\left( \tau \right) d\tau
\right\}\right\rangle &=&\frac{1}{iu\left( t\right)
}\frac{d}{dt}\Theta_{t}\left[ u
\right] \nonumber \\
&=&\frac{\Theta_{t}\left[ u \right]}{iu\left( t\right) }
\frac{d}{dt}\ln \Theta_{t}\left[ u \right] \label{F-2}\\
&=& \frac{\dot \Phi _t\left[ u\right] }{iu\left( t\right) }
\left\langle \exp\left\{i \int\limits_{0}^{t} \xi \left( \tau
\right) u\left( \tau \right) d\tau \right\}\right\rangle .
\nonumber
\end{eqnarray}
Substituting Eq.~(\ref{F-2}) in Eq.~(\ref{F-1}) and using again the
functional translation formula we obtain Klyatskin result
(\ref{Kly}). By using the following integral representation for
$|u|^{\alpha}$
\begin{equation}
|u|^{\alpha} = \frac {\Gamma \left( \alpha +1\right) \sin \left(
\pi \alpha /2\right)}{\pi } \int\limits_{-\infty}^{+\infty}
\frac{1-\cos(xu)}{\left\vert x\right\vert ^{1+\alpha }}dx ,
\nonumber
\end{equation}
we rewrite Eq.~(\ref{Funct}) as
\begin{equation}
\Theta_{t}\left[ u \right] =\exp \left\{ -Q
\int\limits_{0}^{t}d\tau
\int\limits_{-\infty}^{+\infty}\frac{1-\cos\left( xu\left( \tau
\right) \right)}{\left\vert x\right\vert ^{1+\alpha }}dx
\right\},\nonumber
\end{equation}
where
\begin{equation}
Q=\frac{D\Gamma \left( \alpha +1\right) \sin \left( \pi \alpha
/2\right)}{\pi } \; ,  \label{Rel}
\end{equation}
We obtain, therefore, the following expression for the variational
operator in Eq.~(\ref{Kly}) for L\'{e}vy stable noise $L(t)$
\begin{equation}
\frac{\dot{\Phi}_{t}\left[ u\right] }{iu\left( t\right)
}=Q\int\limits_{-\infty}^{+\infty}\frac{e^{ixu\left(
t\right)}-1}{iu\left( t\right) \left\vert x\right\vert ^{1+\alpha
}}\,dx=Q\int\limits_{-\infty}^{+\infty} \frac{dx}{\left\vert
x\right\vert ^{1+\alpha }}\int\limits_{0}^{x}e^{iu\left( t\right)
y}dy.\nonumber
\end{equation}
Substituting this expression in Eq.~(\ref{Kly}) we get
\begin{equation}
\left\langle L\left( t\right) R_{t}\left[ L+z\right] \right\rangle
=Q\int\limits_{-\infty}^{+\infty}\frac{dx}{\left\vert x\right\vert
^{1+\alpha }}\int\limits_{0}^{x}\exp\left\{ y\frac{\delta}{\delta
z\left( t\right) }\right\}\left\langle R_{t}\left[ L+z\right]
\right\rangle dy.\nonumber
\end{equation}
By inserting the operator of functional differentiation into the
average and by putting $z = 0$, we get finally
\begin{equation}
\left\langle L\left( t\right) R_{t}\left[ L\right] \right\rangle =
Q \int\limits_{-\infty}^{+\infty}\frac{dx}{\left\vert x\right\vert
^{1+\alpha }}\int\limits_{0}^{x}\left\langle \exp\left\{
y\frac{\delta}{\delta L\left( t\right) }\right\} R_{t}\left[
L\right] \right\rangle dy.\label{new}
\end{equation}

Now we are ready to derive the fractional Fokker-Planck equation
for L\'{e}vy flights using the functional approach. By
differentiating, with respect to time $t$, the expression for
probability density of random process $x(t)$
\begin{equation}
W\left( x,t\right) =\left\langle \delta \left( x-x\left( t\right)
\right) \right\rangle ,\label{W}
\end{equation}
and taking into account Eq.~(\ref{Lang}), we obtain
\begin{equation}
\frac{\partial W}{\partial t}=\frac{\partial }{\partial x} \,\left[
U^{\prime}(x)W\right] -\frac{\partial }{\partial x}\,\,\left\langle
L\left( t\right) \,\delta \left( x-x\left( t\right) \right)
\right\rangle .  \label{Prelim}
\end{equation}
To evaluate the average in Eq.~(\ref{Prelim}) we apply the formula
(\ref{new})
\begin{equation}
\left\langle L\left( t\right) \,\delta \left( x-x\left( t\right)
\right) \right\rangle = Q \int\limits_{-\infty }^{+\infty
}\frac{dz}{\left\vert z\right\vert ^{1+\alpha
}}\int\limits_{0}^{z}\left\langle \exp \left\{y\,\frac{\delta
}{\delta L\left( t\right) }\right\} \delta \left( x-x\left(
t\right) \right) \right\rangle dy.\label{Fur-Nov}
\end{equation}
Using functional differentiation rules, from Eq.~(\ref{Lang}) we
get
\begin{equation}
\frac{\delta }{\delta L\left( t\right) }\,\delta \left( x-x\left(
t\right) \right) =-\frac{\partial }{\partial x}\,\delta
\left(x-x\left( t\right) \right) \,\frac{\delta x\left( t\right)
}{\delta L\left( t\right) }=-\frac{\partial }{\partial x}\,\delta
\left(x-x\left( t\right) \right) . \label{Equiv}
\end{equation}
Thus, the variational operator $\delta /\delta L\left( t\right) $
with respect to the functional $\delta \left( x-x\left( t\right)
\right) $ is equivalent to the ordinary differential operator
$-\partial /\partial x$. As a result, we have
\begin{equation}
\left\langle L\left( t\right) \,\delta \left( x-x\left( t\right)
\right) \right\rangle =
Q\int\limits_{-\infty}^{+\infty}\frac{dz}{\left\vert z\right\vert
^{1+\alpha }}\,\int\limits_{0}^{z}\exp \left\{ -y\,\frac{\partial
}{\partial x}\right\} \,dy\,W\left( x,t\right) . \label{Split}
\end{equation}
After substitution of Eq.~(\ref{Split}) in Eq.~(\ref{Prelim}) and
evaluation of the internal integral we arrive at
\begin{equation}
\frac{\partial W}{\partial t}=\frac{\partial }{\partial
x}\,\left[U^{\prime }(x)W\right] +Q\int\limits_{-\infty }^{+\infty
}\left[ \exp \left\{-z\, \frac{\partial }{\partial x}\right\}
-1\,\right] \,W\left( x,t\right) \frac{dz}{\left\vert z\right\vert
^{1+\alpha }}\,.\nonumber
\end{equation}
By using the property of the translation operator
\begin{equation}
\exp \left\{ -z\,\frac{d}{dx}\right\} f\left( x\right) =
f\left(x-z\right),\nonumber
\end{equation}
we arrive at the following Kolmogorov equation for the probability
density of nonlinear systems (\ref{Lang}) driven by a symmetric
$\alpha $-stable L\'{e}vy noise
\begin{equation}
\frac{\partial W}{\partial t}=\frac{\partial }{\partial
x}\,\left[U^{\prime }(x)W\right] +Q\int\limits_{-\infty }^{+\infty
}\frac{W\left( x-z,t\right) -W\left( x,t\right) }{\left\vert
z\right\vert ^{1+\alpha }}\,dz. \label{KE}
\end{equation}
The Eq.~(\ref{KE}) represents the well-known Fokker-Planck equation
with fractional space derivative, which describes superdiffusion in
the form of L\'{e}vy flights
\begin{equation}
\frac{\partial W}{\partial t}=\frac{\partial }{\partial x}\,\left[
U^{\prime}(x)W\right] +D\frac{\partial ^{\alpha }W}{\partial
\left\vert x\right\vert ^{\alpha }}\,.  \label{FFPE}
\end{equation}

\section{Stationary probability distributions for L\`{e}vy flights}

First of all, we can try to evaluate the stationary probability
distribution $W_{st}\left( x\right)$ from Eq.~(\ref{FFPE}), if it
exists. Of course, this evaluation is impossible for any potential
profile, but the potential $U(x)$ should satisfy some constraints.
It is better to apply Fourier transform to the
integro-differential equation (\ref{KE}) and to write the equation
for the characteristic function
\begin{equation}
\vartheta \left( k,t\right) =\left\langle e^{ikx\left(
t\right)}\right\rangle =\int\limits_{-\infty }^{+\infty
}e^{ikx}W\left( x,t\right) dx.  \label{Char-0}
\end{equation}
After simple manipulations we find
\begin{equation}
\frac{\partial \vartheta }{\partial t}=-ik\int\limits_{-\infty
}^{+\infty }e^{ikx}U^{\prime }(x)W\left( x,t\right) dx-D\left\vert
k\right\vert ^{\alpha }\vartheta .\,\nonumber
\end{equation}
For smooth potential profiles $U\left( x\right)$, expanding in power
series near the point $x=0$, we can rewrite this equation in the
operator form
\begin{equation}
\frac{\partial \vartheta }{\partial t}=-ikU^{\prime }\left(
-i\frac{\partial }{\partial k}\right) \vartheta -D\left\vert
k\right\vert ^{\alpha }\vartheta .\,  \label{Char-1}
\end{equation}
In particular, for stationary characteristic function, from
Eq.~(\ref {Char-1}) we get
\begin{equation}
U^{\prime }\left( -i\frac{d}{dk}\right) \vartheta _{st}-iD\left\vert
k\right\vert ^{\alpha -1}\mathrm{sgn}\,k\cdot \vartheta _{st}=0\,,
\label{St-Char}
\end{equation}
where $\mathrm{sgn}\,k$ is the sign function. Unfortunately, one
cannot solve Eq.~(\ref{St-Char}) for arbitrary potential $U\left(
x\right) $ and arbitrary L\'{e}vy exponent $\alpha $.

Let us consider, as in \cite{Che02}, the symmetric smooth monostable
potential $U\left( x\right) =\gamma x^{2m}/\left( 2m\right) $
$\left( m=1,2,\ldots \right) $. The Eq.~(\ref{St-Char}), therefore
transforms into the following differential equation of $\left(
2m-1\right) $-order
\begin{equation}
\frac{d^{2m-1}\vartheta _{st}}{dk^{2m-1}}+\left( -1\right)
^{m+1}\beta ^{2m-1}\left\vert k\right\vert ^{\alpha
-1}\mathrm{sgn}\,k\cdot \vartheta _{st}=0\,,  \label{Mon-Pot}
\end{equation}
where $\beta =\sqrt[2m-1]{D/\gamma }$. As it was proved by
analysis of Eq.~(\ref{Mon-Pot}) in~\cite{Che04}, the stationary
probability distribution $W_{st}\left( x\right)$ has non-unimodal
shape and power tails
\begin{equation}
W_{st}\left( x\right) \sim \frac{1}{\left\vert x\right\vert
^{2m+\alpha -1}}\,,\qquad \left\vert x\right\vert \rightarrow
\infty \,.  \label{Tail}
\end{equation}
In Ref.~\cite{Che04}, the estimation of bifurcation time for
transition from unimodal initial distribution to bimodal
stationary one and the existence of a transient trimodal state for
$m>2$ were found.

Exact solution of Eq.~(\ref{Mon-Pot}) can be only obtained for the
case of Cauchy noise: $\alpha =1$. Due to the symmetry of the
characteristic function $\vartheta _{st}\left( -k\right) =\vartheta
_{st}\left( k\right) $ we can reduce Eq.~(\ref{Mon-Pot}) to the
linear differential equation with constant parameters
\begin{equation}
\frac{d^{2m-1}\vartheta _{st}}{dk^{2m-1}}-\left( -1\right) ^{m}\beta
^{2m-1}\vartheta _{st}=0\qquad \left( k>0\right) .  \label{Cauchy}
\end{equation}
From the corresponding characteristic equation
\begin{equation}
\lambda ^{2m-1}=\left( -1\right) ^{m}\beta ^{2m-1},
\end{equation}
we select the roots with negative real part, which are meaningful
from physical point of view. The general solution of
Eq.~(\ref{Cauchy}), therefore, reads
\begin{eqnarray}
&&\vartheta _{st}\left( k\right) =\sum_{l=0}^{\left[ \left(
m-1\right) /2 \right] }A_{l}\exp \left\{ -\beta \left\vert
k\right\vert \cos
\frac{\pi \left( m-2l-1\right) }{2m-1}\right\}\cdot \nonumber\\
&&\cos \left( \beta \left\vert k\right\vert \sin \frac{\pi \left(
m-2l-1\right) }{2m-1}-\varphi _{l}\right) ,  \label{General}
\end{eqnarray}
where the quadratic brackets in the upper limit of the sum denote
the integer part of the expression. The unknown constants $A_{l}$
and $\varphi _{l}$ can be calculated from the obvious conditions
\begin{equation}
\vartheta _{st}\left( 0\right) =1,\quad \vartheta _{st}^{\left(
2j-1\right) }\left( +0\right) =0\quad \left( j=1,2,\ldots
,m-1\right) .\label{Cond}
\end{equation}
Substituting Eq.~(\ref{General}) in Eq.~(\ref{Cond}) we have
\begin{eqnarray}
&&\sum_{l=0}^{\left[ \left( m-1\right) /2\right] }A_{l}\cos
\varphi _{l}=1, \label{Cond-2}\\
&&\sum_{l=0}^{\left[ \left( m-1\right) /2\right] }A_{l}\cos \left[
\frac{\pi \left( 2j-1\right) \left( m+2l\right) }{2m-1}-\varphi
_{l}\right] =0\quad \left( j=1,2,\ldots ,m-1\right) .\nonumber
\end{eqnarray}
Making the reverse Fourier transform of Eq.~(\ref{General}) we
obtain the stationary probability distribution (SPD) of the particle
coordinate
\begin{equation}
W_{st}\left( x\right) = \frac{\beta }{\pi }\sum_{l=0}^{\left[
\left( m-1\right) /2\right] }A_{l}\frac{x^{2}\cos \left[ \frac{\pi
\left( m-2l-1\right) }{2m-1}+\varphi _{l}\right] +\beta ^{2}\cos
\left[ \frac{\pi \left( m-2l-1\right) }{2m-1}-\varphi _{l}\right]
}{x^{4}-2x^{2}\beta ^{2}\cos \frac {\pi \left( 4l+1\right) }{2m-1}
+\beta ^{4}}\,. \label{SPD}
\end{equation}

The parabolic potential profile $U\left( x\right) = \gamma x^{2}/2$
corresponds to a linear system (\ref{Lang}). In this situation, from
Eqs.~(\ref{Cond-2}) and (\ref{SPD}) we easily obtain the following
obvious result
\begin{equation}
W_{st}\left( x\right) =\frac{\beta }{\pi \left( x^{2}+\beta
^{2}\right) }\,, \label{Linear}
\end{equation}
\ie due to the stability of the Cauchy distribution
(\ref{Linear}), the probabilistic characteristics of driving noise
increments (see Eq.~(\ref{Ch-func})) and Markovian process
$x\left(t\right)$ are similar.

For quartic potential $\left( m=2\right)$, from the set of
Eq.~(\ref{Cond-2}), we find $A_{0}=2/\sqrt{3}$, $\varphi _{0}=\pi
/6$. Substituting these parameters in Eq.~(\ref{SPD}) we obtain
\begin{equation}
W_{st}\left( x\right) =\frac{\beta ^{3}}{\pi \left( x^{4}-x^{2}\beta
^{2}+\beta ^{4}\right) }\,,  \label{Quartic}
\end{equation}
which coincides, for $\beta = 1$, with the result obtained in
Ref.~\cite {Che02}. The plots of stationary probability
distributions (\ref{Quartic}) for L\'{e}vy flights in symmetric
quartic potential for different values of parameter $ \beta $ are
shown in Fig.~1.
\begin{figure}[tbph]
\begin{center}
\includegraphics[width=7cm]{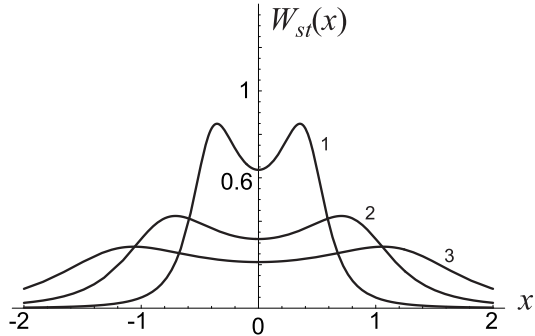}
\end{center}
\caption{{\protect\small \emph{Stationary probability distributions for
L\'{e}vy flights in symmetric quartic potential $U(x)=\protect\gamma %
x^{4}/4$ for different values of dimensionless parameter
$\protect\beta $: $1$ - $\protect\beta =0.5$, $2$ - $\protect\beta
=1$, $3$ - $\protect\beta =1.5$.}\protect\bigskip }}
\label{beta_1}
\end{figure}
The superdiffusion in the form of L\'{e}vy flight gives rise to a
bimodal stationary probability distribution when the particle
moves in a monostable potential, differently from the ordinary
diffusion of the Brownian motion characterized by unimodal SPD.

The SPD of superdiffusion has two maxima at the points $x=\pm
\beta /\sqrt{2}$, with the value $\left( W_{st}\right) _{max}
=4/\left( 3\pi \beta \right) $. Since the value of the minimum is
$W_{st}\left( 0\right) =1/\left( \pi \beta \right) $, the ratio
between maximum and minimum value is constant and equal to $4/3$.
The width of probability density increases with increasing
parameter $\beta =\sqrt[3]{D/\gamma }$, \ie with decreasing the
steepness $\gamma $ of the quartic potential profile, or with
increasing the noise intensity $D$.

Carrying out analogous procedure we obtain the stationary
probability distributions for the cases $m=3,4,5$
\begin{eqnarray}
W_{st}\left( x\right) &=& \frac{\beta ^{5}}{\pi \left( x^{2}+\beta
^{2}\right) \left( x^{4}-2\beta ^{2}x^{2}\cos \pi /5+\beta
^{4}\right)}\,, \nonumber\\
W_{st}\left( x\right) &=& \frac{\beta ^{7}}{\pi \left(
x^{4}-2\beta ^{2}x^{2}\cos \pi /7+\beta ^{4}\right) \left(
x^{4}+2\beta ^{2}x^{2}\cos 2\pi /7+\beta ^{4}\right) }\,, \label{Distr}\\
W_{st}\left( x\right) &=& \frac{\beta ^{9}}{\pi \left( x^{2}+\beta
^{2}\right) \left( x^{4}-2\beta ^{2}x^{2}\cos \pi /9+\beta
^{4}\right) \left( x^{4}+2\beta ^{2}x^{2}\cos 4\pi /9+\beta
^{4}\right) }\,. \nonumber
\end{eqnarray}
The plots of distributions (\ref{Distr}), for different values of
parameter $\beta $, are respectively shown in Figs.~2--4. It must
be emphasized that according to Figs.~2--4,
\begin{figure}[tbph]
\begin{center}
\includegraphics[width=7cm]{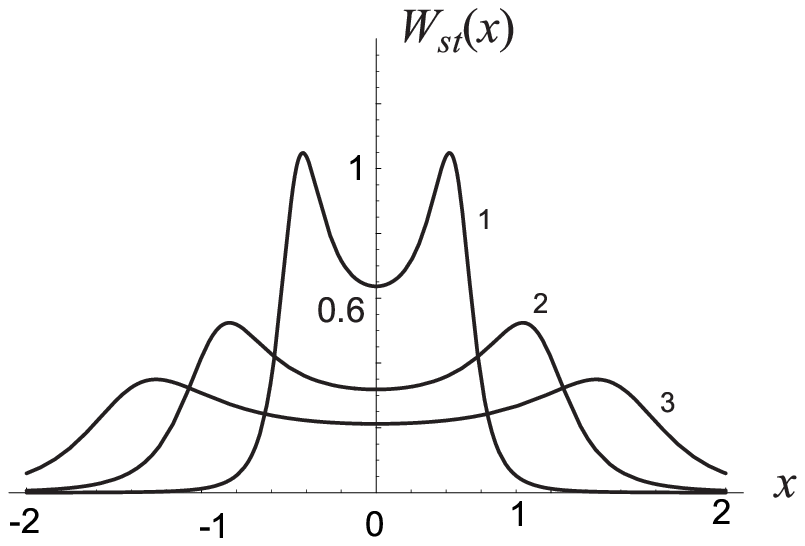}
\end{center}
\caption{{\protect\small \emph{Stationary probability
distributions for L\'{e}vy flights in symmetric potential
$U(x)=\protect\gamma x^{6}/6$ for different values of
dimensionless parameter $\protect\beta $: $1$ - $\protect\beta
=0.5$, $2$ - $\protect\beta =1$, $3$ - $\protect\beta =1.5$.}}}
\end{figure}
\begin{figure}[tbph]
\begin{center}
\includegraphics[width=7cm]{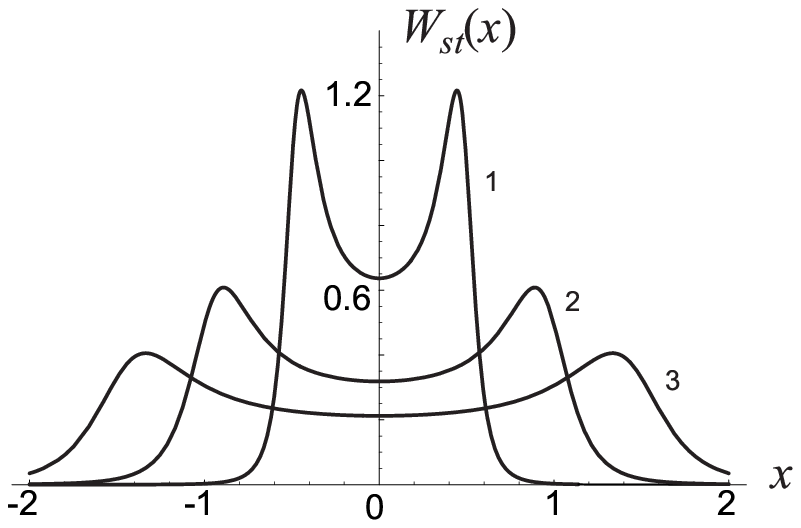}
\end{center}
\caption{{\protect\small \emph{Stationary probability
distributions for L\'{e}vy flights in symmetric potential
$U(x)=\protect\gamma x^{8}/8$ for different values of
dimensionless parameter $\protect\beta $: $1$ - $\protect\beta
=0.5$, $2$ - $\protect\beta =1$, $3$ - $\protect\beta =1.5$.}}}
\end{figure}
\begin{figure}[tbph]
\begin{center}
\includegraphics[width=7cm]{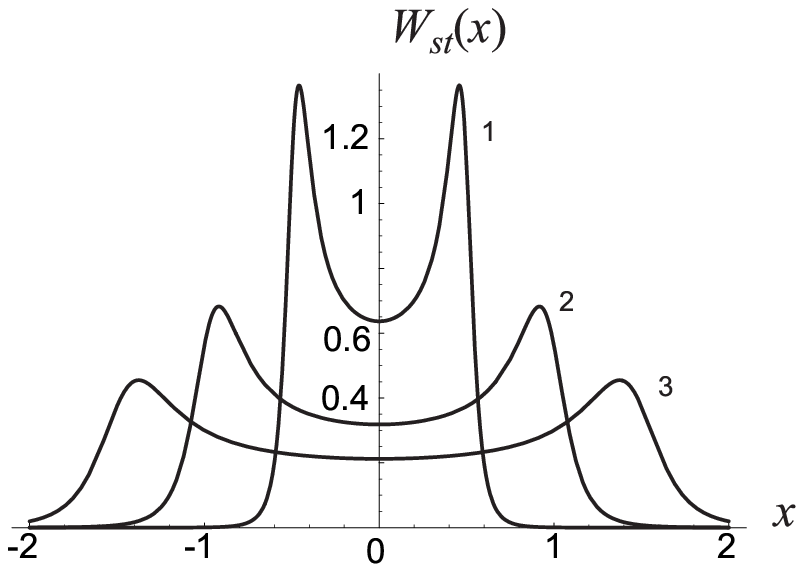}
\end{center}
\caption{{\protect\small \emph{Stationary probability
distributions for L\'{e}vy flights in symmetric potential
$U(x)=\protect\gamma x^{10}/10$ for different values of
dimensionless parameter $\protect\beta $: $1$ - $\protect\beta
=0.5$, $2$ - $\protect\beta =1$, $3$ -
$\protect\beta =1.5$.}%
\protect\bigskip }} \label{beta_4}
\end{figure}
these distributions remain bimodal and have the same tendency with
increasing $\beta $, but the ratio between maximum and minimum
increases with increasing $m$. From Eqs.~(\ref{Quartic}) and
(\ref{Distr}) we see that the second moment of the particle
coordinate is finite for $m \geq 2$. This means that there is a
confinement of the particle motion due to the steep potential
profile, even if the particle moves according to a superdiffusion
in the form of L\'{e}vy flights~\cite{Che04}. The presence of two
maxima is a peculiarity of the superdiffusion motion. Because of
the fast diffusion due to L\'{e}vy flights, the particle reaches
very quickly regions near the potential walls on the left or on
the right with respect to the origin $x = 0$. Then the particle
diffuses around this position, until a new flight moves it in the
opposite direction to reach the other potential wall. As a result,
the particle spends a large time in some symmetric areas with
respect to the point $x=0$, differently from the Brownian
diffusion in monostable potential profiles. These symmetric areas
lie near the maxima of the bimodal SPD. For fixed $D$ and $m$,
these maxima are closer or far away the point $x = 0$ depending on
the greater or smaller steepness $\gamma$ of the potential
profile. This corresponds to a greater or smaller confinement of
the particle motion. Of course, such a confinement is more
pronounced for greater $m$, that is for steeper potential
profiles.

On the basis of Eqs.~(\ref{Linear})--(\ref{Distr}) and the known
behavior of density tails (\ref{Tail}), we can write the general
expressions for stationary probability distribution in the case of
potential $U\left( x\right) =\gamma x^{2m}/\left( 2m\right) $ with
odd $m=2n+1$
\begin{equation}
W_{st}\left( x\right) =\frac{\beta ^{4n+1}}{\pi \left( x^{2}+\beta
^{2}\right) }\prod\limits_{l=0}^{n-1}\frac{1}{x^{4}-2\beta
^{2}x^{2}\cos \left[ \pi \left( 4l+1\right) /\left( 4n+1\right)
\right] +\beta ^{4}}\,,\label{Final-1}
\end{equation}
and even $m=2n$
\begin{equation}
W_{st}\left( x\right) =\frac{\beta ^{4n-1}}{\pi
}\prod\limits_{l=0}^{n-1} \frac{1}{x^{4}-2\beta ^{2}x^{2}\cos
\left[ \pi \left( 4l+1\right) /\left( 4n-1\right) \right] +\beta
^{4}}\, , \label{Final-2}
\end{equation}
which are, together with Eqs.~(\ref{Quartic})-(\ref{Distr}), the
main result of this paper.

\section{Conclusions}

We used functional analysis approach to derive the fractional
Fokker-Planck equation directly from Langevin equation with
symmetric $\alpha $-stable L\'{e}vy noise. This approach allows to
describe anomalous diffusion in the form of L\'{e}vy flights. We
obtained the general formula for stationary probability distribution
of superdiffusion in symmetric smooth monostable potential for
Cauchy driving noise. All distributions have bimodal shape and
become more narrow with increasing steepness of the potential or
with decreasing noise intensity. We found that the variance of the
particle coordinate is finite for quartic potential profile and for
steeper potential profiles, that is a confinement of the particle in
a superdiffusion motion in the form of L\'{e}vy flights. As a
result, we can evaluate the power spectral density of a stationary
motion. Calculations of residence times for the case of L\'{e}vy
flights in bistable potential with steep potential wells, and
anomalous diffusion in periodic ratchet-like potentials are the
subjects of forthcoming investigations.

This work has been supported by MIUR, CNISM, and by Russian
Foundation for Basic Research (project 05-02-16405).

\end{document}